\documentclass[apj]{emulateapj}

\usepackage{amsmath}
\usepackage{txfonts}

\input{epsf}

\begin{document}
\def\lax    {\ifmmode{_<\atop^{\sim}}\else{${_<\atop^{\sim}}$}\fi}
\def\gax    {\ifmmode{_>\atop^{\sim}}\else{${_>\atop^{\sim}}$}\fi}
\def\gtorder{\mathrel{\raise.3ex\hbox{$>$}\mkern-14mu
             \lower0.6ex\hbox{$\sim$}}}
\def\ltorder{\mathrel{\raise.3ex\hbox{$<$}\mkern-14mu
             \lower0.6ex\hbox{$\sim$}}}
 
\long\def\***#1{{\sc #1}}

\title{Bright X-ray Transients in M31: 2004 July {\em XMM-Newton} Observations.}

\author{Sergey Trudolyubov\altaffilmark{1,2}, William Priedhorsky\altaffilmark{3}, 
and France Cordova\altaffilmark{1}}

\altaffiltext{1}{Institute of Geophysics and Planetary Physics, University of 
California, Riverside, CA 92521}
\email{sergeyt@ucr.edu}

\altaffiltext{2}{Space Research Institute, Russian Academy of Sciences, 
Profsoyuznaya 84/32, Moscow, 117810 Russia}

\altaffiltext{3}{Los Alamos National Laboratory, Los Alamos, NM 87545}

\begin{abstract}
We present the results of X-ray observations of four bright transients sources 
detected in the July 2004 {\em XMM-Newton} observations of the central bulge of 
M31. Two X-ray sources, XMMU J004315.5+412440 and XMMU J004144.7+411110, were 
discovered for the first time. Two other sources, CXOM31 J004309.9+412332 and 
CXOM31 J004241.8+411635, were previously detected by {\em Chandra}. The properties 
of the sources suggest their identification with accreting binary systems in M31. 
The X-ray spectra and variability of two sources, XMMU J004144.7+411110 and 
CXOM31 J004241.8+411635, are similar to that of the Galactic black hole transients, 
making them a good black hole candidates. The X-ray source XMMU J004315.5+412440 
demonstrates a dramatic decline of the X-ray flux on a time scale of three days, 
and a remarkable flaring behavior on a short time scales. The X-ray data on 
XMMU J004315.5+412440 and CXOM31 J004309.9+412332 suggest that they can be either 
black hole or neutron star systems. Combining the results of 2000-2004 {\em XMM} 
observations of M31, we estimate a total rate of the bright transient outbursts in 
the central region of M31 to be 6 - 12 yr$^{-1}$, in agreement with previous 
studies. 
\end{abstract}


\section{INTRODUCTION}
Bright X-ray transient sources provide a unique opportunity to study the properties 
of the accretion onto stellar-mass compact objects. The observations of the Galactic 
X-ray novae (XRNe) show that they can be either high-mass (HMXB) and low-mass (LMXB) 
binaries \cite{TS96}, but most of confirmed black-hole binaries are low-mass systems 
(McClintock \& Remillard 2005, and references therein). 

Until recently, the detailed study of X-ray transient sources has been mostly limited 
to our Galaxy. The advent of a new generation of X-ray telescopes ({\em Chandra} and 
{\em XMM-Newton}) has allowed the study of the spectral and temporal properties of XRNe 
located in the nearby galaxies and compare them to their Galactic counterparts. 

The relative proximity and favorable orientation of M31 make it a prime target for 
the study of an extragalactic XRNe population. Recent {\em Chandra} and {\em XMM-Newton} 
observations of M31 have led to the discovery of several dozen bright transient X-ray 
sources with luminosities between $\sim 10^{36}$ and $\sim 10^{38}$ ergs s$^{-1}$ (Garcia 
et al. 2000; Trudolyubov et al. 2001; Kong et al. 2002; Williams et al. 2004; Williams et 
al. 2005{\em a, b, c}). The follow-up observations of X-ray transients with {\em HST} 
ensured the identification and study of optical counterparts to some of these systems 
(Williams et al. 2004; Williams et al. 2005{\em a, b}). 

Here we present the results of spectral and timing analysis of four bright transient 
sources detected in the 2004 July {\em XMM-Newton} observations.

\section{OBSERVATIONS AND DATA ANALYSIS}
\begin{figure}
\epsfxsize=16cm
\epsffile{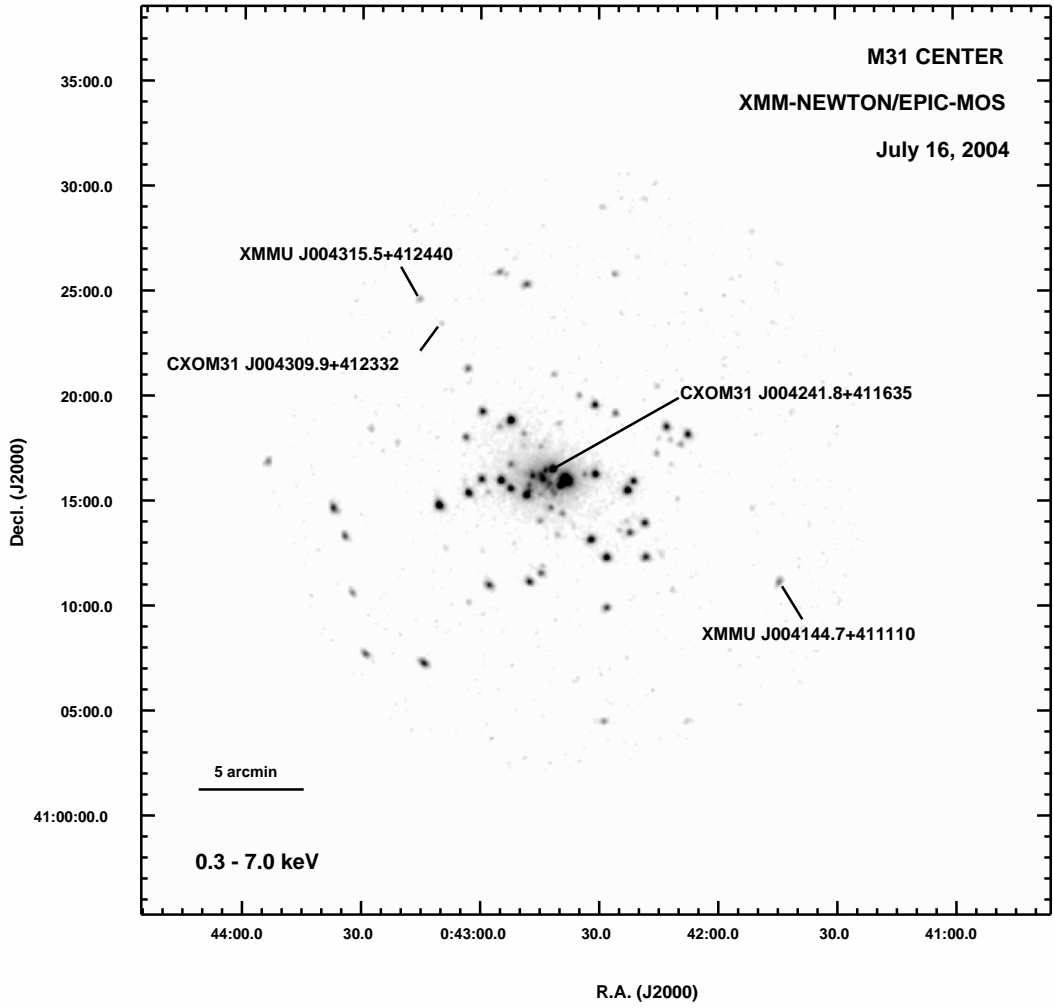}
\caption{The central region of M31 as it appears in {\em XMM-Newton} EPIC-MOS 2004 
July 16 observation. The transient sources are marked with arrows.
\label{image_general}}
\end{figure}

The central region of M31 was observed with {\em XMM-Newton} on four occasions during 
July 2004 \cite{Barnard_Bo158}(Fig. \ref{image_general}). In the following analysis 
we use the data from three European Photon Imaging Camera (EPIC) instruments: two EPIC 
MOS detectors \cite{Turner01} and the EPIC-pn detector \cite{Strueder01}. In all 
observations EPIC instruments were operated in the {\em full window} mode ($30\arcmin$ 
FOV) with the {\em medium} optical blocking filter.

We reduced EPIC data with the latest version of {\em XMM-Newton} Science Analysis System 
(SAS v 6.5.0)\footnote{See http://xmm.vilspa.esa.es/user}. Each of the original event files 
were screened for periods of high background. The remaining exposure times for each 
observation are listed in Table \ref{obslog}. The 2004, July 17 observation (Obs. $\# 2$ 
in Table \ref{obslog}) is affected by high background, so we excluded it from our spectral 
and timing analysis, and used it for total flux estimates only.

We generated EPIC-pn and MOS images of the central region of M31 (Fig. \ref{image_general}) 
in the 0.3 -- 7.0 keV energy band, and used the SAS standard maximum likelihood (ML) source 
detection script {\em edetect\_chain} to detect and localize point sources. We used bright 
X-ray sources with known optical counterparts from USNO-B \cite{Monet03} and 2MASS catalogs 
\cite{Cutri03} to correct EPIC image astrometry. After applying the astrometric correction, we 
estimate residual systematic error in the source positions to be of the order $0.5 - 1\arcsec$. 
To identify transient sources, the resulting list of the {\em XMM} sources was compared to the 
existing catalogs of M31 X-ray sources (Trinchieri \& Fabbiano 1991; Primini et al. 1993; Kong 
et al. 2002; Williams et al. 2004; Pietsch et al. 2005) and transient source lists from 
the {\em Chandra} monitoring campaign (e.g. Williams et al. 2005{\em a, b, c}). 

To generate lightcurves and spectra of X-ray sources, we used elliptical extraction regions 
with semi-axes size of $\sim 15 - 50 \arcsec$ (depending on the distance of the source from 
the telescope axis) and subtracted as background the spectrum of adjacent source-free regions, 
with subsequent normalization by ratio of the detector areas. For spectral analysis, we used 
data in the $0.3 - 7$ keV energy band. All fluxes and luminosities derived from spectral 
analysis apply to this band. We used spectral response files generated by XMM SAS tasks. Spectra 
were grouped to contain a minimum of 20 counts per spectral bin in order to allow $\chi^{2}$ 
statistics and fit to analytic models using the 
XSPEC v.11\footnote{http://heasarc.gsfc.nasa.gov/docs/xanadu/xspec/index.html} fitting package 
\cite{arnaud96}. EPIC-pn, MOS1 and MOS2 data were fitted simultaneously, but with normalizations 
varying independently. The energy spectra of the sources were fitted by two standard X-ray binary 
spectral models \cite{MR04}: absorbed simple power law or multicolor disk blackbody models (DISKBB). 
To estimate upper limits on the quiescent source luminosities, the EPIC count 
rates were converted into energy fluxes in the $0.3 - 7$ keV energy band using Web 
PIMMS\footnote{See http://heasarc.gsfc.nasa.gov/Tools/w3pimms.html}, assuming 
standard parameters: an absorbed simple power law model with $N_{\rm H} = 7 \times 10^{20}$ cm$^{-2}$ 
(Galactic foreground value \cite{DL90}) and photon index $\alpha = 1.7$ \cite{Shirey01}. We used 
standard XANADU/XRONOS v.5\footnote{http://heasarc.gsfc.nasa.gov/docs/xanadu/xronos/xronos.html} 
tasks to perform analysis of the timing properties of the transient X-ray sources.

In the following analysis we assume M31 distance of 760 kpc (van den Bergh 2000). All parameter 
errors quoted are $68\%$ ($1\sigma$) confidence limits.

\section{RESULTS AND DISCUSSION}

\subsection{XMMU J004144.7+411110}
A new X-ray source XMMU J004144.7+411110 has been discovered in the data of 2004 July 16 
{\em XMM-Newton} observations of M31. In addition, our analysis of the archival {\em Chandra} 
data (2004, July 17 observation $\# 4719$) revealed the presence of a bright X-ray source 
at the position consistent with XMMU J004144.7+411110. Combining the data of {\em XMM-Newton} 
and {\em Chandra} observations, we measure the position of XMMU J004144.7+411110 to be 
$\alpha = 00^{h} 41^{m} 44.70^{s}, \delta = 41\arcdeg 11\arcmin 10\arcsec$ 
(J2000 equinox) with an uncertainty of $\sim 1\arcsec$. The search for the optical counterparts 
using the images from Local Group Survey (LGS) \cite{Massey01} did not yield any object 
brighter than $m_{\rm v} \sim 21$ within the error circle of XMMU J004144.7+411110. Using the 
data of archival {\em XMM} observations of the central region of M31, we estimate the upper 
limit ($2\sigma$) on the source quiescent luminosity to be $\sim 2\times 10^{35}$ ergs s$^{-1}$ 
in the $0.3 - 7$ keV energy band, $>100$ times lower than maximum measured outburst 
luminosity (Table \ref{spec_par}).

\begin{figure}
\epsfxsize=16.0cm
\epsffile{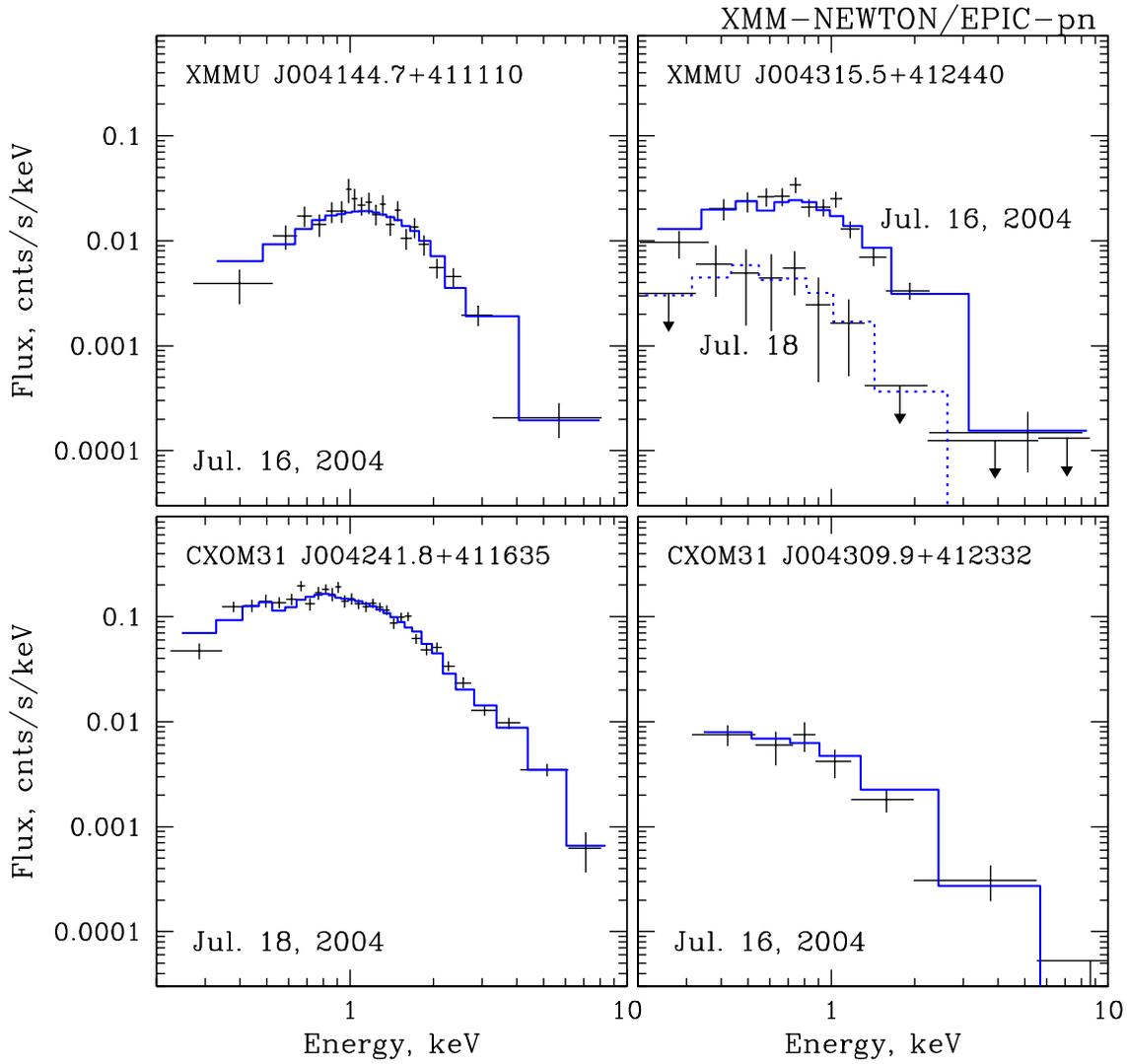}
\caption{\small Count spectra of bright transient sources. EPIC-pn data, $0.3 - 7$ keV 
energy range. The analytical model fits are shown with {\em thick histograms}. {\em Upper left:} 
Transient source XMMU J004144.7+411110 (absorbed DISKBB model). {\em Upper right:} XMMU 
J004315.5+412440 (absorbed power law model). An absorbed power law model fit to the data of 
July 18 observation is shown with {\em dotted} histogram. {\em Lower left:} CXOM31 J004241.8+411635 
(absorbed DISKBB+power law model). {\em Lower right:} CXOM31 J004309.9+412332 (absorbed power law 
model). 
\label{spec_TR_fig}}
\end{figure}

The energy spectra of XMMU J004144.7+411110 are soft, and can be well fitted by absorbed DISKBB 
model with color temperatures $\sim 0.6 - 0.8$ keV or absorbed power law model with photon index 
of $\sim 2.8 - 3.3$ (Table \ref{spec_par}; Fig. \ref{spec_TR_fig}). The corresponding estimated 
luminosities of the source have been found to be in the range of $\sim (2.3 - 3.0)\times 10^{37}$ 
ergs s$^{-1}$. For two observations (1 and 3), the energy spectrum shows clear signs of high 
energy cut-off: the DISKBB model approximates it better than a simple power law, as indicated by 
fit statistics (Table \ref{spec_par}). 

The X-ray spectrum, transient behavior and extreme faintness of the optical counterpart indicate 
that XMMU J004144.7+411110 is not a Galactic foreground object, and probably is an accreting 
binary system in M31. The spectral model fits require absorbing columns well in excess the Galactic 
foreground value of $7 \times 10^{20}$ cm$^{-2}$ (Table \ref{spec_par}), this could be consistent 
with the source located inside or behind the M31 disk \cite{TP04}.     

The observed spectrum and luminosity of XMMU J004144.7+411110 bear clear resemblance to the Galactic 
black-hole transients in the {\em high}/``thermal-dominant'' state during the flux decline that 
precedes the transition to the {\em low}/{\em hard} state (Tomsick \& Kaaret 2000; McClintock \& 
Remillard 2005). It should be also noted, that the $0.3 -7$ keV spectrum of the source is significantly 
softer than observed in the neutron star systems at similar luminosity levels both in the Galaxy 
\cite{CS97} and in M31 globular clusters \cite{TP04}. 

\subsection{XMMU J004315.5+412440}
2004 July {\em XMM} observations revealed another previously undetected X-ray source, located 
at $\alpha = 00^{h} 43^{m} 15.51^{s}, \delta = 41\arcdeg 24\arcmin 40\arcsec \pm 1.5\arcsec$. 
The inspection of the LGS images showed no optical sources brighter than $m_{\rm v} \sim 21$ in 
the {\em XMM} error circle of XMMU J004315.5+412440. Using the data of previous {\em XMM} 
observations, we estimated quiescent luminosity of the source to be $\lesssim 10^{35}$ ergs s$^{-1}$ 
in the $0.3 - 7$ keV energy band.

\begin{figure}
\epsfxsize=16.0cm
\epsffile{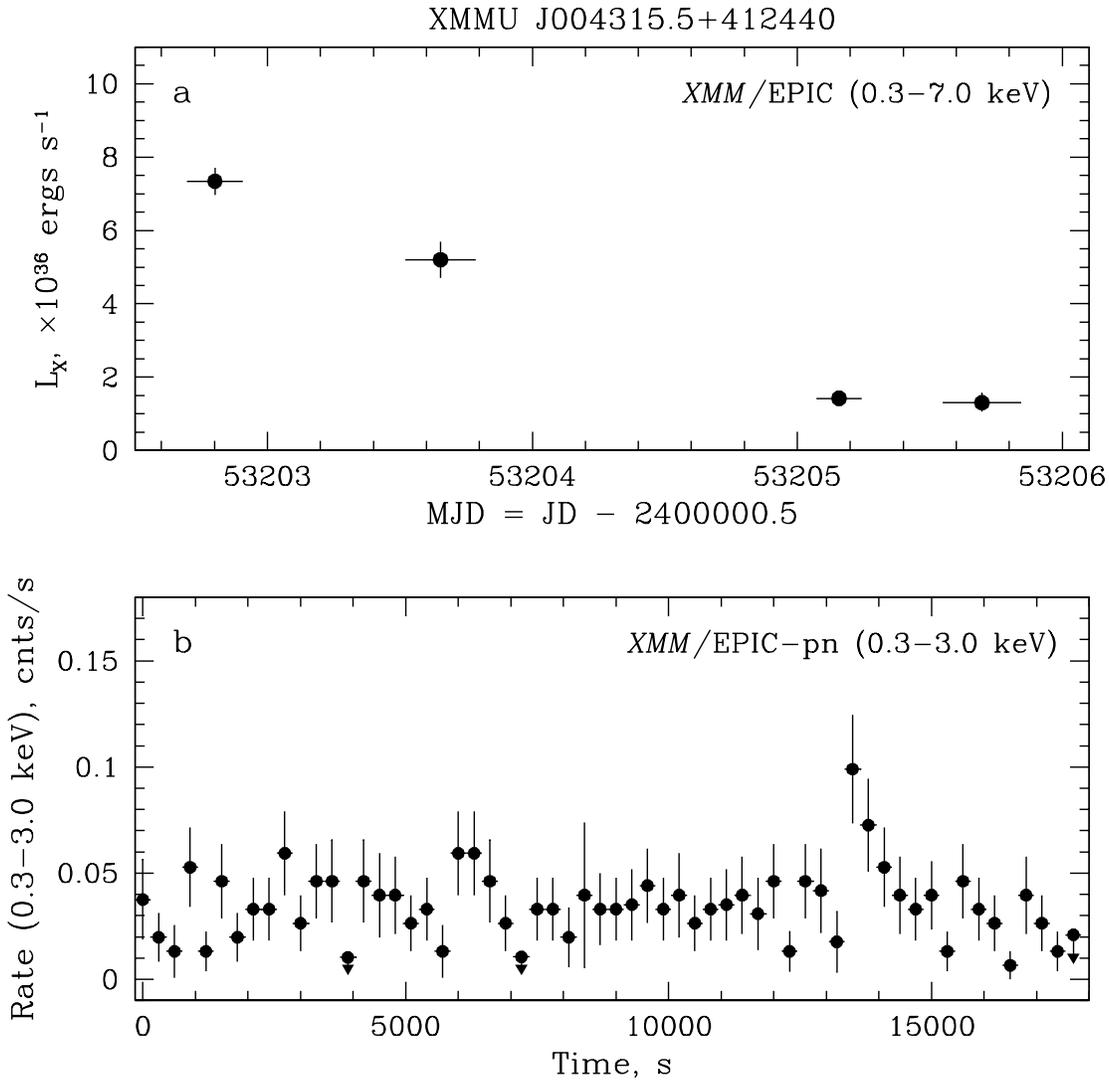}
\caption{\small {\em a:} The long-term X-ray light curve of the transient source 
XMMU J004315.5+412440 during four 2004 July {\em XMM-Newton} observations, obtained 
from combined data of EPIC-pn, MOS1 and MOS2 cameras in the $0.3 - 7.0$ keV energy 
range. {\em b:} Background and vignetting corrected X-ray light curve of XMMU 
J004315.5+412440 during 2004 July 16 {\em XMM-Newton} observation (Obs. $\# 1$), 
obtained from data of EPIC-pn in the $0.3 - 3.0$ keV energy range, with a 300 s 
time resolution. 
\label{lc_c_125_fig}}
\end{figure}

The X-ray source XMMU J004315.5+412440 demonstrates a remarkable variability on time scales ranging 
from minutes to several days (Fig. \ref{lc_c_125_fig}). The source flux in the $0.3 - 7$ keV band 
changed dramatically in the course of four 2004 July {\em XMM} observations, dropping from 
$\sim 10^{-13}$ erg s$^{-1}$ on July 16 to $\sim 1.5\times 10^{-14}$ erg s$^{-1}$ in four days 
(Fig. \ref{lc_c_125_fig}, {\em upper panel}). The source XMMU J004315.5+412440 also shows a high 
level of variability during July 16 observation (Obs. 1)(Fig. \ref{lc_c_125_fig}, {\em lower panel}). 
In addition to the irregular flux variations, the source produced an intense flare at $\sim 13700$ s 
(Fig. \ref{lc_c_125_fig}, {\em lower panel}), lasting for $\sim 1000$ s. The time evolution of the 
source flux during the flare is characterized by fast rise ($< 200$ s) to a maximum level followed 
by quasi-exponential decay with estimated e-folding time of $\sim 500\pm200$ s. The peak intensity 
of the flare was $> 4$ times higher than the average source intensity, corresponding to a luminosity 
of $\sim 3\times 10^{37}$ erg s$^{-1}$ in the $0.3 - 7$ keV energy band. The estimated total energy 
emitted during the flare is $\sim 10^{40}$ ergs. Unfortunately, the sensitivity of our observations 
does not allow to make a reliable conclusion on the evolution of the source spectrum during this flare.

The X-ray spectrum of XMMU J004315.5+412440 during July 16 observation (Obs. 1) is soft, and can 
be well approximated by an absorbed  power law model with photon index of $\sim 3.8$ or by DISKBB 
model with color temperature of $\sim 0.36$ keV (Table \ref{spec_par}). The corresponding absorbed 
luminosity of the source was $\sim 7\times 10^{36}$ erg s$^{-1}$. The power law model provides better 
fit to the observational data than a DISKBB model, as seen from Table \ref{spec_par}. The power law 
model fit requires a high level of low-energy absorption $\sim 5\times 10^{21}$ cm$^{-2}$, while the 
DISKBB model requires an absorbing column consistent with Galactic foreground value in the direction 
of M31 (Table \ref{spec_par}). We did not detect a statistically significant change in the shape of 
the source spectrum during the overall flux decline, as seen from power law model fits 
(Table \ref{spec_par}). 

The X-ray properties of XMMU J004315.5+412440 along with the constraints on the optical counterpart 
support its identification as an accreting binary system in M31. A combination of strong long and 
short-term variability of X-ray flux makes XMMU J004315.5+412440 especially interesting. The profile 
and luminosity of the X-ray flare detected in the July 16 observations are somewhat similar to the 
long thermonuclear X-ray bursts detected from the Galactic neutron stars, although the spectrum of 
the source is significantly softer than that of the burst sources \cite{SB05}. The other possibility 
is that the flare is caused by spasmodic accretion onto the compact object, similar to the neutron 
star Type II bursts \cite{Lewin95}. 

\subsection{CXOM31 J004241.8+411635}
The X-ray source CXOM31 J004241.8+411635 was first detected in the July 17, 2004 {\em Chandra} 
observation and remained detectable in the 2004 Sept. 2 and Oct. 4 {\em Chandra} observations 
\cite{Williams05_2}. The source has been previously detected in 1979 {\em Einstein} observations 
\cite{TF91}, which makes it a probable recurrent transient candidate with duty cycle of 
$0.02 - 0.06$ \cite{Williams05_2}. In addition, the follow-up observations of the source with 
{\em HST} resulted in a determination that the optical counterpart to CXOM31 J004241.8+411635 
is a low-mass binary system in M31 \cite{Williams05_2}.

The X-ray spectra of CXOM31 J004241.8+411635 measured with {\em XMM}/EPIC are relatively soft (Fig. 
\ref{spec_TR_fig}), and can be approximated by absorbed power law with photon index of $\sim 2.3 - 2.5$, 
or by DISKBB model with characteristic temperature of $\sim 0.86 - 0.99$ keV (Table \ref{spec_par}). 
Due to the curvature in the source spectrum around $1$ keV, the DISKBB model provides significantly 
better description to the data than a power law. The power law model fits result in a relatively high 
values of the absorbing column ($N_{\rm H} \sim 2\times 10^{21}$ cm$^{-2}$), while in the DISKBB 
approximation, the measured absorption was consistent with Galactic foreground absorption toward M31. 
The X-ray spectra can be also fitted by combination of the DISKBB and power law models with 
$kT_{in} \sim 0.9 - 1.1$ keV and a photon index of $\sim 4 - 5$, improving the overall quality of the 
fit with respect to one-component model approximation.

The estimated luminosity of the source changes from $5\times 10^{37}$ to $\sim 6.5\times 10^{37}$ 
ergs s$^{-1}$ in the course of {\em XMM} observations. There is a clear correlation between the 
spectral temperature derived in the DISKBB spectral fits and the X-ray flux (Table \ref{spec_par}), 
similar to the results of the {\em Chandra} observations of the source \cite{Williams05_2}. At the 
same time, the characteristic emitting radius $R_{in} \sqrt{cos~i}$ remains essentially constant, 
despite a significant change in the X-ray flux (Table \ref{spec_par}), the effect also observed in 
the Galactic black hole candidates in the {\em high} spectral state \cite{TL95}. 

The X-ray properties of CXOM31 J004241.8+411635 observed with {\em XMM} have been found to be 
remarkably similar to the properties of the Galactic black hole candidates in the {\em high} 
or ``thermal-dominant'' state \cite{MR04}, in general agreement with the results of {\em Chandra} 
observations \cite{Williams05_2}, which makes it a good black hole candidate.   
    
\subsection{CXOM31 J004309.9+412332}
The X-ray source CXOM31 J004241.8+411635 was first detected in the 2004 May 23 {\em Chandra} 
observation \cite{Williams05_1}. Regular {\em Chandra} monitoring of the source revealed a complex 
X-ray light curve with at least two peaks separated by $\sim 130$ days and reaching the maximum 
$0.3 - 7$ keV luminosity of $\sim 10^{37}$ ergs s$^{-1}$. Based on the results of {\em Chandra} and 
{\em HST} observations, CXOM31 J004309.9+412332 has been classified as a low-mass X-ray binary 
system \cite{Williams05_1}.

The {\em XMM-Newton} spectra of CXOM31 J004241.8+411635 during 2004 July {\em XMM} observations can 
be approximated by steep absorbed power law with photon index of $\sim 2.8 - 4.0$ or by DISKBB model 
with $kT_{in} \sim 0.33 - 0.36$ keV (Table \ref{spec_par}). The corresponding absorbed luminosity of 
the source is in the range of $\sim (2 - 3) \times 10^{36}$ ergs s$^{-1}$. The spectra of CXOM31 
J004241.8+411635 measured with {\em XMM-Newton}, correspond to the decline of the first outburst, 
and appear to be significantly softer than measured during 2004, May 23 and Sep. 2 {\em Chandra} 
observations (coincident with two outburst peaks) \cite{Williams05_1}. The anti-correlation of the 
spectral hardness and X-ray flux of CXOM31 J004241.8+411635 could be similar to that observed in 
some Galactic low-mass X-ray binaries (both neutron star and black hole binary systems) and in 
globular cluster sources in M31 \cite{TP04}. The available data on CXOM31 J004241.8+411635 suggest 
that it could be either a black hole or a neutron star, not allowing to make a definitive conclusion 
on the nature of the compact object in this system.

\section{Conclusions}
Using the data of 2004 July {\em XMM}/EPIC observations of M31, we study the X-ray properties of 
four bright transient sources. Two X-ray sources, XMMU J004315.5+412440 and XMMU J004144.7+411110, 
were discovered for the first time. Two other sources, CXOM31 J004309.9+412332 and CXOM31 
J004241.8+411635, were previously detected by {\em Chandra}. The properties of the sources along 
with the information on optical counterparts suggest their identification with accreting binary 
systems belonging to M31. The luminosities, energy spectra and variability of the two sources, XMMU 
J004144.7+411110 and CXOM31 J004241.8+411635, are reminiscent of the Galactic black hole X-ray novae, 
which makes them a good black hole candidates. The X-ray source XMMU J004315.5+412440 demonstrates a 
dramatic decline of the X-ray flux on a time scale of three days, and a remarkable flaring behavior 
during the first {\em XMM} observation. The available data on this source and the other transient 
source CXOM31 J004309.9+412332 is consistent with either black hole or a neutron star interpretation.

The total of 10 transient sources with $0.3 - 7$ keV luminosities higher that $10^{36}$ ergs s$^{-1}$ 
have been detected in 5 {\em XMM-Newton} observations of the central part of M31 (Osborne et al. 2001; 
Trudolyubov et al. 2001; Pietsch et al. 2005; this Letter). More than a half of these sources ($60\%$) 
can be classified as black hole candidates. The remaining sources include two supersoft transients with 
probable white dwarf primaries ($20\%$), and two systems, which can contain either black holes or 
neutron stars. 

Given a number of the detected sources, and the coverage factor of the {\em XMM} observations, one can 
estimate the expected total rate of X-ray transient outbursts in the bulge and inner disk of M31. 
Assuming the average duration of a bright phase of the typical transient of $\sim 1 - 2$ months, gives 
a total rate of $\sim 6 - 12$ outbursts per year\footnote{Note a large uncertainty of this estimate, due 
to the limited statistics and scatter in the observed outburst durations.}, consistent with estimates based 
on earlier {\em XMM-Newton} and {\em Chandra} results (Trudolyubov et al. 2001; Kong et al. 2002; Williams 
et al. 2004).

\section{Acknowledgments}
Support for this work was provided through NASA Grant NAG5-12390. XMM-Newton is an ESA Science Mission 
with instruments and contributions directly funded by ESA Member states and the USA (NASA). This research 
has made use of data obtained through the High Energy Astrophysics Science Archive Research Center Online 
Service, provided by the NASA/Goddard Space Flight Center.

\begin{table}
\small
\caption{2004 July {\em XMM-Newton} observations of the central region of M31. 
\label{obslog}}
\small
\begin{tabular}{ccccccc}
\hline
\hline
Obs. $\#$ &Date, UT & $T_{\rm start}$, UT & Obs. ID  & RA (J2000)$^{a}$ & 
Dec (J2000)$^{a}$ & Exp.(pn)$^{b}$\\
& &(h:m:s)&&  (h:m:s)   &(d:m:s)&(ks)\\             
\hline
$\#1$    &2004 Jul 16    &16:17:05&0202230201&00:42:42.12&41:16:57.1& 16 \\
$\#2^{c}$&2004 Jul 17    &12:07:53&0202230401&00:42:42.08&41:16:56.9& 18 \\
$\#3$    &2004 Jul 18-19 &23:49:30&0202230401&00:42:42.26&41:16:58.2& 13 \\
$\#4$    &2004 Jul 19    &12:48:23&0202230501&00:42:42.23&41:16:57.6& 6.5 \\
\hline
\end{tabular}
\begin{list}{}{}
\item[$^{a}$] -- coordinates of the center of the field of view
\item[$^{b}$] -- instrument exposure used in the analysis
\item[$^{c}$] -- observation affected by high background
\end{list}
\end{table}

\begin{table}
\small
\caption{Model fits to the energy spectra of transient sources ({\em XMM}/EPIC data, 
$0.3 - 7$ keV energy range). \label{spec_par}}
\small
\begin{tabular}{cccccccccc}
\hline
\hline
 Obs. &Model & N$_{\rm H}$                &kT   &$R_{in} \sqrt{cos~i}^{a}$ &Photon&Flux$^{b}$&$\chi^{2}$&$L_{\rm X}^{c}$&Instrument\\
 $\#$ &      &($\times 10^{20}$ cm$^{-2}$)&(keV)&    (km)           &Index &          &(d.o.f)   & &          \\      
\hline
\hline
\multicolumn{9}{c}{XMMU J004144.7+411110}\\
\hline
1&  PL  &$51\pm5$&             &        &$3.16\pm0.15$         &$4.38\pm0.16$&  $56.6(53)$& 30.3 &pn+M1+M2\\ 
 &DISKBB&$19\pm3$&$0.68\pm0.04$&$30\pm4$&                      &$4.32\pm0.16$&  $48.8(53)$& 29.9 &\\
\hline
3&  PL  &$40^{+5}_{-4}$&       &        &$2.72^{+0.14}_{-0.12}$&$3.59\pm0.16$&  $64.8(53)$& 24.8 &pn+M1+M2\\
 &DISKBB&$14\pm3$&$0.78^{+0.06}_{-0.05}$&$20^{+3}_{-4}$&       &$3.36\pm0.15$&  $67.6(53)$& 23.2 &\\
\hline
4&  PL  &$55\pm7$&             &        &$3.31^{+0.12}_{-0.22}$&$3.89\pm0.20$&  $54.9(44)$& 26.9 &pn+M2\\         
 &DISKBB&$22\pm4$&$0.60\pm0.06$&$37^{+8}_{-7}$&                &$3.66\pm0.19$&  $51.0(44)$& 25.3 &\\
\hline
\hline
\multicolumn{9}{c}{XMMU J004315.5+412440}\\
\hline
1&  PL  &$29\pm5$&             &        &$3.81^{+0.44}_{-0.32}$&$1.07\pm0.06$&  $33.9(22)$& 7.39 &pn\\
 &DISKBB&$7^{+4}_{-3}$&$0.36^{+0.05}_{-0.04}$&$51\pm14$&       &$0.98\pm0.06$&  $38.7(22)$& 6.77 &\\
\hline
3&  PL  &$18^{+16}_{-18}$&     &        &$4.10^{+2.44}_{-1.67}$&$0.20\pm0.03$&    $4.1(5)$& 1.38 &pn\\
\hline
\hline
\multicolumn{9}{c}{CXOM31 J004241.8+411635}\\
\hline
1&  PL  &$23\pm2$&             &        &$2.54^{+0.05}_{-0.06}$&$7.26\pm0.13$&$220.8(209)$& 50.2 &pn+M1+M2\\
 &DISKBB&$3\pm2$&$0.86\pm0.03$&$20\pm2$ &                      &$7.07\pm0.13$&$198.1(209)$& 48.9 &        \\
\hline
3&  PL  &$21\pm1$&             &        &$2.36^{+0.04}_{-0.05}$&$8.07\pm0.12$&$295.8(233)$& 55.8 &pn+M1+M2\\
 &DISKBB&$3\pm1$&$0.94\pm0.03$&$18\pm1$&                       &$7.77\pm0.12$&$284.4(233)$& 53.7 &\\
\hline
4&  PL  &$22\pm1$&             &        &$2.33\pm0.04$         &$9.65\pm0.21$&$300.8(242)$& 66.7 &pn+M1+M2\\
 &DISKBB&$4\pm1$&$0.99^{+0.02}_{-0.03}$ &$18\pm1$&             &$9.50\pm0.21$&$263.9(242)$& 65.7 &\\
\hline
\hline
\multicolumn{9}{c}{CXOM31 J004309.9+412332}\\
\hline
1&  PL  &$11^{+8}_{-4}$&        &        &$2.80\pm0.50$&$0.38\pm0.05$&$11.9(8)$& 2.63 &pn\\
 &DISKBB&$4^{+11}_{-3}$&$0.33\pm0.11$    &$33^{+84}_{-18}$&&$0.28\pm0.04$&$15.8(8)$& 1.94 &\\
\hline
3&  PL  &$31^{+12}_{-6}$&      &        &$4.07^{+1.75}_{-0.36}$&$0.44\pm0.05$&$13.4(11)$& 3.04 &pn\\
 &DISKBB&$6^{+6}_{-1}$&$0.36^{+0.08}_{-0.07}$&$34^{+28}_{-14}$&&$0.43\pm0.05$&$14.1(11)$& 2.97 &\\
\hline 
\end{tabular}

\begin{list}{}{}
\item $^{a}$ -- effective inner disk radius, where $i$ is the inclination angle of the disk 
\item $^{b}$ -- absorbed model flux in the $0.3 - 7$ keV energy range in 
units of $10^{-13}$ erg s$^{-1}$ cm$^{-2}$
\item $^{c}$ -- absorbed luminosity in the $0.3 - 7$ keV energy range in 
units of $10^{36}$ erg s$^{-1}$, assuming the distance of 760 kpc
\end{list}
\end{table}

\end{document}